\documentclass[twocolumn]{aastex61}
\title{C/2017 K2 Paper}

\newcommand\aastex{AAS\TeX}

\usepackage{xcolor}
\usepackage{url}

\received{September 21, 2017}
\revised{October 4, 2017}
\accepted{October 6, 2017}

\submitjournal{ApJ Letters}

\shorttitle{\aastex\ C/2017 K2 PANSTARRS}
\shortauthors{Meech et al.}

\begin{document}

\title{CO-Driven Activity in Comet C/2017 K2 (PANSTARRS)}

\correspondingauthor{Karen J. Meech}
\email{meech@ifa.hawaii.edu}

\author[0000-0002-2058-5670]{Karen J. Meech}
\affiliation{Institute for Astronomy \\
2680 Woodlawn Drive \\
Honolulu, HI 96822 USA}

\author[0000-0002-4734-8878]{Jan T. Kleyna}
\affiliation{Institute for Astronomy \\
2680 Woodlawn Drive \\
Honolulu, HI 96822 USA}

\author[0000-0001-6952-9349]{Olivier Hainaut}
\affiliation{European Southern Observatory \\
Karl-Schwarzschild-Strasse 2\\
D-85748 Garching bei M\"unchen, Germany}

\author[0000-0001-7895-8209]{Marco Micheli}
\affiliation{ESA SSA-NEO Coordination Centre \\
Largo Galileo Galilei, 1 \\
00044 Frascati (RM), Italy}
\affiliation{INAF - Osservatorio Astronomico di Roma \\
Via Frascati, 33 \\
00040 Monte Porzio Catone (RM), Italy}

\author[0000-0003-1790-5749]{James Bauer}
\affiliation{University of Maryland, Dept. of Astronomy \\
College Park, MD 20742-2421 USA}

\author[0000-0002-7034-148X]{Larry Denneau}
\affiliation{Institute for Astronomy \\
2680 Woodlawn Drive \\
Honolulu, HI 96822 USA}

\author[0000-0002-2021-1863]{Jacqueline V. Keane}
\affiliation{Institute for Astronomy \\
2680 Woodlawn Drive \\
Honolulu, HI 96822 USA}

\author{Haynes Stephens}
\affiliation{University of California at Berkeley, 501 Campbell Hall \\
Berkeley, CA 94720 USA}

\author[0000-0001-7830-028X]{Robert Jedicke}
\affiliation{Institute for Astronomy \\
2680 Woodlawn Drive \\
Honolulu, HI 96822 USA}

\author[0000-0002-1341-0952]{Richard Wainscoat}
\affiliation{Institute for Astronomy \\
2680 Woodlawn Drive \\
Honolulu, HI 96822 USA}

\author[0000-0002-0439-9341]{Robert Weryk}
\affiliation{Institute for Astronomy \\
2680 Woodlawn Drive \\
Honolulu, HI 96822 USA}

\author[0000-0002-1050-4056]{Heather Flewelling}
\affiliation{Institute for Astronomy \\
2680 Woodlawn Drive \\
Honolulu, HI 96822 USA}

\author[0000-0002-7696-0302]{Eva Lilly}
\affiliation{Institute for Astronomy \\
2680 Woodlawn Drive \\
Honolulu, HI 96822 USA}

\author[0000-0002-7965-2815]{Eugene Magnier}
\affiliation{Institute for Astronomy \\
2680 Woodlawn Drive \\
Honolulu, HI 96822 USA}

\author[0000-0001-6965-7789]{Kenneth C. Chambers}
\affiliation{Institute for Astronomy \\
2680 Woodlawn Drive \\
Honolulu, HI 96822 USA}

\begin{abstract}

Comet C/2017 K2 (PANSTARRS) was discovered by the Pan-STARRS1 (PS1) Survey on 2017 May 21 at  a distance 16.09 au from the Sun, the second most distant discovery of an active comet.  Pre-discovery images in the PS1 archive back to 2014 and additional deep CFHT images between 2013 May 10-13 showed the comet to be active at 23.75 au.  We derive an upper limit to the nucleus radius of $R_N$=80 km, assuming a 4\% albedo. The spectral reflectivity of the comet surface is similar to ``fresh'' regions seen on comet 67P/Churyumov-Gerasimenko using the $Rosetta$ OSIRIS camera.  Pre-discovery photometry combined with new data obtained with Megacam on the CFHT show that the activity is consistent with CO-ice sublimation and inconsistent with CO$_2$-ice sublimation.  The ice sublimation models were run out to perihelion in 2022 at 1.8 au to predict the CO production rates, assuming that the outgassing area does not change.  Assuming a canonical 4\% active surface area for water-ice sublimation, we present production rate ratios, $Q_{\rm CO}$/$Q_{\rm H2O}$, for a range of nucleus sizes. Comparing these results with other CO-rich comets we derive a lower limit to the nucleus radius of $\sim$14 km.  We present predictions for $Q_{\rm CO}$ at a range of distances that will be useful for planning observations with JWST and large ground-based facilities.

\end{abstract}



\section{Introduction} 
\label{sec:intro}

Comet C/2017 K2 is a dynamically new Oort cloud comet on a hyperbolic orbit ($e$=1.0007) that was discovered on 2017 May 21 by the Pan-STARRS1 telescope.  At magnitude 20.8 the comet was at $r=16.09$~au, $\Delta = 16.02$~au and at a true anomaly (TA) of $-140.8^\circ$ moving towards perihelion at $q$ = 1.81 au, which the comet will reach on 2022 Dec. 21.6.  

To compare to the discovery distances for other long period (LP) comets, we searched the Minor Planet Center comet database \texttt{CmtObs.dat}\footnote{\url{http://www.minorplanetcenter.net/iau/ECS/MPCAT-OBS/MPCAT-OBS.html}} of all long-period comet observations back to 1950 (see Fig.~\ref{fig:discovery}).  We assumed that the discovery is the date of the earliest observation. For some comets, the earliest MPC observations might be pre-discovery recoveries (precoveries), so comets at $r>10$ au at their first observation were investigated further to remove precovery observations. Then we used OpenOrb \citep{granvik2009} to compute the heliocentric distance at time of discovery. Only C/2010 U3 (Boattini) was discovered farther from the Sun, at 18.4 au. Thus, C/2017 K2 is the second most distant discovery of an active comet, and pre-discovery observations at $r=23.75$ au represent the largest distance at which an active comet has been observed approaching perihelion.  Most of the historically bright LP comets discovered prior to 1950 were discovered much closer to the sun, with only a few exceptions at $r>5$ au, and none were discovered inbound at distances greater than 6.5 au \citep{roemer1962}.  There were a few bright historical comets for which dust-dynamical models suggest activity began as far out as 30 au \citep{sekanina1975}.

The proliferation of all-sky surveys such as LINEAR \citep{stokes2000}, Spacewatch \citep{gehrels1996}, the Catalina Sky Survey (CSS) \citep{larson2007}, LONEOS \citep{bowell1995}, and NEAT \citep{pravdo1999} in the mid-1990s, followed by Pan-STARRS1 \citep{kaiser2010,chambers2017} in 2010 has resulted in a rapid increase in the discovery of faint active comets at increasingly large heliocentric distances (Fig.~\ref{fig:discovery}). Inbound comets, being heated for the first time, provide unique insights into the mechanisms of comet activity.
 
\section{Observations and Data Reduction} 
\label{sec:observations}

\begin{figure*}[ht!]
\plotone{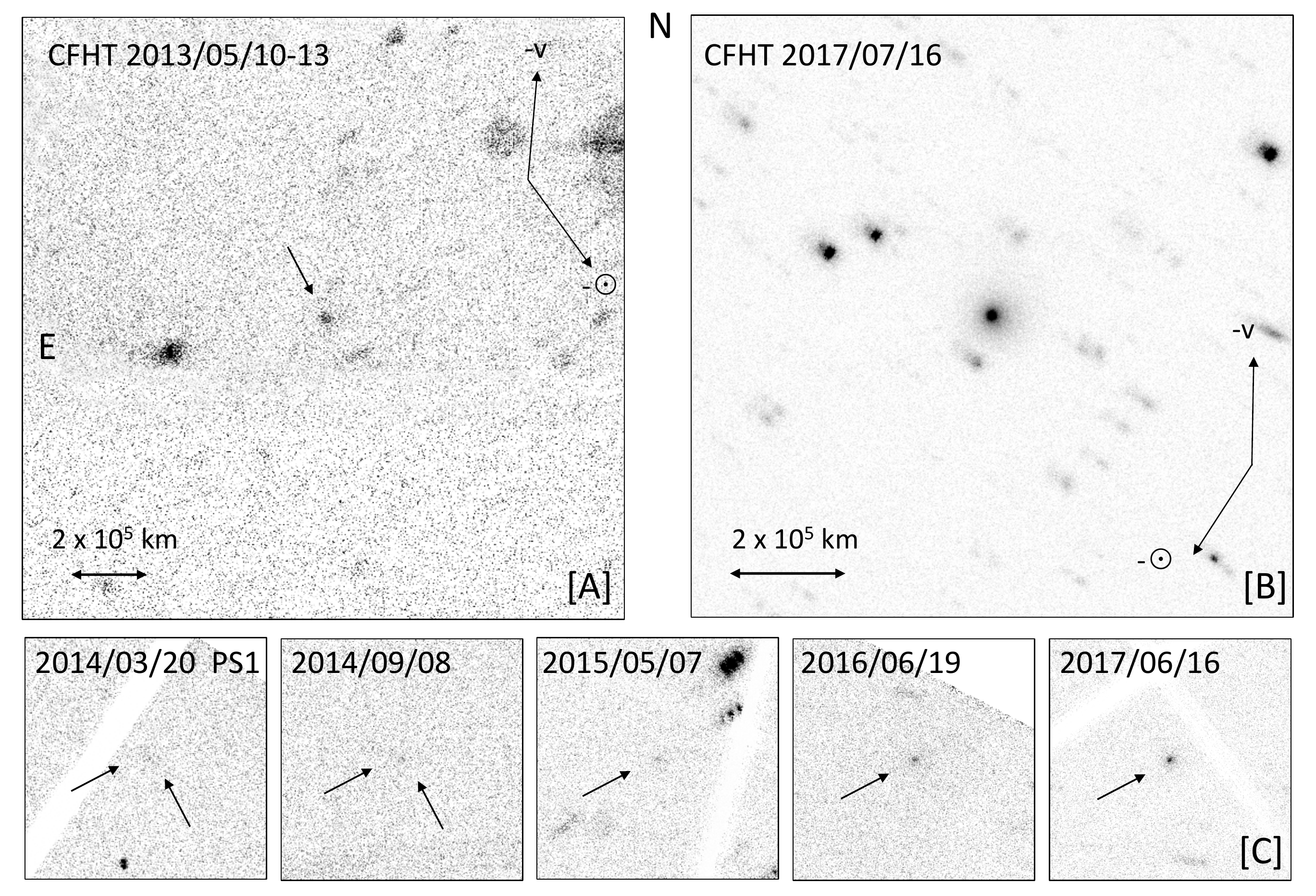}
\caption{Composite images of C/2017 K2. [A] This 4200 sec composite (100$\times$100$''$) used the 7 best u-band CFHT images from 2013 May obtained from the CADC archive.  [B] 2017 July 16 CFHT $r$-band composite with total exposure 720 sec.  The image is 100$''$ on a side.  The coma extends $\sim$3.5$\times$10$^5$ km toward the W.  The negative of the heliocentric velocity (-v) and the extended Sun-target vector or anti-solar direction are shown. [C] Composite images from the PS1 archive for several dates.  All are $i$-band, except the first, which is $r$-band.  All images are 1$\times$$10^6$ km on a side. 
\label{fig:image}}
\end{figure*}

Photometry for C/2017 K2 was obtained using both the CFHT and Pan-STARRS1 (PS1) telescopes. The headers were used to download orbital elements from the Minor Planet Center, and the computed object location was used to determine which object in the frame corresponded to the target.  Terapix tools (SExtractor) were used to produce multi-aperture and automatic aperture target photometry. To photometrically calibrate both telescopes we calculated a photometric zero point for each image using the Pan-STARRS database and published color corrections to translate photometric bands \citep{magnier2017,chambers2017}.  Photometry and observing circumstances are presented in Table~\ref{tab:data}, and a selection of images is shown in Fig.~\ref{fig:image}.

\subsection{Pan-STARRS1}
\label{sec:PS1}

A search for pre-discovery observations in the PS1 images taken between 2010 to 2017, resulted in almost 200 images at the comet's location. The comet is visible in about half of the frames after 2014, while older images or images in narrower passbands are not deep enough to detect it. However, the astrometry from the positive detections constrains the ephemeris to less than one pixel over the entire arc.  This allowed us to measure a lower magnitude limit in images where it was not visible.

We measured the photometry for these PS1 images using a 2$\farcs$5 radius aperture.  When the comet was too faint and SExtractor was unable to locate it, the photometry was done by placing an aperture at the comet's expected position.  The data reported in Table~\ref{tab:data} represent the weighted average magnitudes from all detections on a given night. Conversions to the SDSS photometric system used the transformations from \citet{tonry2012}. For images where  SExtractor was unable to locate the comet, but it was visible to the observer, we confirmed that the measurements were of the comet by inspection.  In all cases the comet appeared extended. We measured the curve of growth for frames where the comet was visible at high $S/N$ to estimate an aperture correction of $\Delta m$ = -0.63 mag to convert to a 5$''$ radius uniform aperture for comparison of all the data to the models in $\S$\ref{sec:models}. To obtain limiting magnitudes where the comet was not visible, we plotted on each field the mean-magnitudes of the stars from the PS1 PV3 catalog (which utilizes the best data reduction and calibration). The limiting magnitude in each field, at which stars were no longer visible, was at S/N$\approx 2$. Using only the observations available on the MPC website, the uncertainty in the orbit position (the long semi-major axis of the 1-$\sigma$ uncertainty ellipse) is $\pm$0$\farcs$15 for the entire period from 2010 to 2017.  Including the new pre-discovery data found in the PS1 images, the error is even smaller for some periods.  With a plate scale of 0$\farcs$25 per pixel, the positional uncertainty is $<$ 1 pixel in the images. For most frames the limiting magnitude was around $r$$\sim$$i$$\sim$21.3.

\vspace{0.2cm}
\subsection{CFHT}
\label{sec:CFHT}

We obtained additional images using the CFHT MegaCam wide-field imager, an array of forty 2048$\times$4612 pixel CCDs with a plate scale of 0$\farcs$187 per pixel and a 1.1 square degree FOV.  The data were obtained through SDSS filters using queue service observing and were processed to remove the instrumental signature through the Elixir pipeline \citep{magnier2004}.  The colors for this active LP comet are shown in Table~\ref{tab:data} and are consistent with other active LP comets \citep{jewitt2015}.  We have converted the colors to a relative spectral reflectivity using 

\begin{equation}
R_{\lambda} = \frac{10^{-0.4(m_{\lambda}-m_{\lambda\odot})}}
	{10^{-0.4(m_o -m_{o \odot})}} = \frac{N}{D}
\end{equation}

\begin{equation}
	\sigma_{R\lambda} = R_{\lambda}^2 
	\left[\left(\frac{0.9212 N \sigma_\lambda}{N}\right)^2
	    + \left(\frac{0.9212 D \sigma_{m o}} {D}\right)^2\right]^{0.5}
\end{equation}

\noindent
Here $N$ and $D$ represent the numerator and denominator in Eq. (1), $m_{\lambda}$ is the magnitude in a specific filter $\lambda$, $\sigma_\lambda$ is the uncertainty on $m_\lambda$, $m_o$ is the reference bandpass that we normalize to, and $m_{\odot}$ is the absolute magnitude of the sun. For the SDSS filters we use $g_{\odot}$ = 5.12$\pm$0.02, $r_{\odot}$ = 4.69$\pm$0.03, $i_{\odot}$ = 4.57$\pm$0.03, and $z_{\odot}$ = 4.60$\pm$0.03\footnote{\url{http://www.sdss.org/dr12/algorithms/ugrizvegasun/}}. We normalized the spectral reflectivities to $\lambda$=0.65 $\mu$m.  The spectral reflectivity is shown in Fig.~\ref{fig:spectrum} in comparison with the reflectivity from several regions from the surface of comet 67P/Churyumov-Gerasimenko as imaged by the $Rosetta$ OSIRIS instrument \citep{fornasier2017}.

We used the Solar System Object Image Search tool at the Canadian Astronomy Data Centre \citep{gwyn2012} to search all archival data stored there for images that might have had pre-discovery detections of the comet.  Eleven $u$-band exposures were found from 2013 May 10-13 obtained with Megacam on the CFHT for a total integration time of 6600 sec. These images were also processed by the Elixir pipeline. The comet is clearly visible at the expected position and appears diffuse. The magnitude of the comet was measured on the combination of the best 7 images giving $u=23.09\pm0.17$. We used our measured ($u-r$)=2.333$\pm$0.055 color index to convert to $r$=20.76$\pm$0.23  (Table~\ref{tab:data}).

\subsection{NEOWISE}
\label{sec:NEOWISE}

The NEOWISE survey \citep{mainzer2014} observed C/2017 K2 during two visits. The first was for 76 exposures between 2017-03-27 17:35:44.627 UT and 2017-04-06 07:02:15.385 UT, with a mid-frame observing time of 2017-04-01 20:10:26.526 UT. The second visit was for 65 exposures between 2017-06-27 01:38:28.719 UT and 2017-07-08 20:27:56.365 UT with a mid-frame observing time of 2017-07-02 23:03:12.423 UT. The mid-frame heliocentric distances were $r$=15.8 au and $r$=16.4 au, respectively. The $W2$ band encompasses both the CO 1-0 and CO$_2$ $\nu_3$ emission bands.  Because the ratio of the CO$_2$ to CO $g$-factors is $\sim$11.2 \citep{bockelee1989}, a given flux implies a much higher production rate for CO than CO$_2$.  These visits showed no significant detections, and using the techniques described in \citet{bauer2015} yielded 3-$\sigma$ upper CO production rate limits of $Q_{CO} <$ 1.6$\times$10$^{28}$ and $Q_{CO} <$ 1.0$\times$10$^{28}$ molecules per second respectively, and upper CO$_2$ production limits of $Q_{CO2} <$ 1.4$\times$10$^{27}$ and $Q_{CO2} <$ 8.9$\times$10$^{26}$ molecules per second respectively. 

\section{Analysis}
\label{sec:analysis}

\subsection{Sublimation Models}
\label{sec:models}

We used a surface ice sublimation model \citep{meech1986} to investigate the activity for comet C/2017 K2.  The model computes the amount of gas sublimating from an icy surface exposed to solar heating, as described in detail in \citet{meech2017}.  The total brightness within a fixed aperture combines radiation scattered from both the nucleus and the dust dragged from the nucleus in the escaping gas flow, assuming a dust to gas mass ratio of 1. This type of model can distinguish between H$_2$O, CO, and CO$_2$ driven activity.  The model free parameters include: nucleus radius, albedo, emissivity, nucleus density, dust properties, and fractional active area.  When there is information about some of the parameters, it is possible to constrain many of the others.

Because C/2017 K2 is a recent discovery, none of the model parameters are constrained.  However, based on typical values for other comets seen in-situ and from the ground \citep{meech2017b}, we assumed the following:  nucleus albedo, $p_v$=0.04, emissivity, $\epsilon$=0.9, nucleus phase function, $\beta$=0.04 mag deg$^{-1}$, coma phase function, $\beta_c$=0.02 mag deg$^{-1}$, and nucleus density, $\rho_N$=400 kg m$^{-3}$, and an average dust size of 2 $\mu$m.  With steep power law size distributions for grains ranging in size between 0.1$\mu$m-mm, the small particles dominate \citep{fulle2016}. The grain sizes can't be modeled using a dust-dynamical techniques because this comet has severe projection effects (i.e. we are looking straight down the tail). With knowledge of the nucleus size, the fractional active area can be fit. However, as this comet was discovered active, we have no a priori knowledge of the nucleus size thus the only parameter we can constrain is the effective surface area of the sublimating ice for each volatile.  

We ran a suite of models with a range of nucleus sizes and found that for a radius $R_N$=80 km the model brightness (nucleus+coma) approached that of the photometry--unrealistic given that all of the images showed a dust coma, indicating activity.  We thus use $R_N$=80 km as an upper limit to the nucleus size assuming an albedo $p_v$=0.04.


Assuming that the CO outgassing surface area remains constant, with no CO$_2$ contribution through perihelion, and that the fractional nucleus surface area for H$_2$O-ice sublimation is 4\%-typical of other nuclei without icy halos \citep{ahearn1995}, we can infer the minimum nucleus radius.  Data for the $\sim$30 comets with high-quality simultaneous H$_2$O and CO production rate ($Q$) measurements show that at perihelion and within 2 au, when water-sublimation is strong, the ratios of Q$_{CO}$/Q$_{H2O}$ are below 30\% \citep{paganini2014, meech2017b}.  Fitting models with a range of nucleus sizes to calculate the production rates for H$_2$O and CO at perihelion, we rule out nuclei with $R_N<$14 km because they would require Q$_{CO}$/Q$_{H2O}\ge$30\% and an active surface fraction of 0.092\% for CO sublimation. This suggests that the C/2017 K2 nucleus with $14<R_N<80$ km could be as large as C/1995 O1 (Hale-Bopp) which had $R_N\sim30\pm10 $ km \citep{fernandez2002}.

Figure~\ref{fig:model}A shows the best fit model for a nucleus radius of 14 km for sublimation from CO or CO$_2$, forced to match the photometric data at the time of discovery at TA=-140.8$^{\circ}$.  We also show a fit for an 80 km radius nucleus. The data from 2017 May to September show insufficient range along the orbit (TA) to distinguish between sublimation from CO or CO$_2$ as a driver of the activity.  At these distances, there is no contribution from H$_2$O sublimation.  However, the pre-discovery data from PS1, and the archival CFHT data show very clearly that only the CO-sublimation model can reproduce the photometry. Changing the nucleus size increases or decreases the nucleus contribution to the total brightness, but at these distances has no effect on the shape of the light curve.  We ran models for CO and CO$_2$ sublimation that produced gas flows consistent with the gas production rate limits obtained from the WISE data (see $\S$~\ref{sec:NEOWISE}) and have plotted the corresponding expected limits on the total brightness in the figure. According to the best fit models, the maximum grain size that could be lifted off at these distances for CO$_2$ sublimation is $\sim$2 $\mu$m, and for CO sublimation a few 100 $\mu$m. The dust grain size we used for the models is well below this limit.

\section{Discussion}
\label{sec:discuss}

The PS1 limiting magnitudes at $\sim$30 au and the precovery data until discovery are consistent with a steady sublimation from the surface. The model is brighter than the limit at TA=-149$^{\circ}$, but this could reflect a lower comet brightness possibly due to nucleus rotation. The difference is not significant enough to interpret this as a sublimation decrease.

There are several possible mechanisms for activity at these large distances. The equilibrium sublimation temperatures of the most abundant ices that can drive activity, CO, CO$_2$ and H$_2$O, are 25K, 80K and 160K, respectively.  Sublimation rate is a non-linear function of temperature, and can occur at low rates at large distances.  The distance at which surface-ice sublimation becomes effective at driving comet activity is when the gas flow lifts sufficient dust from the surface to be detected from Earth. For water this is within the distance of Jupiter; for CO$_2$, between Saturn and Uranus; and for CO, within the Kuiper belt \citep{meech2009}.  Volatiles condensing below 100K can also be trapped in amorphous water ice and their release occurs as the ice is heated and undergoes restructuring through annealing or the amorphous-to-crystalline ice transition.  This transition begins around 120K and annealing begins at temperatures as low as 37K.  CO is the only abundant cometary volatile that can reproduce the C/2017 K2 lightcurve shape from sublimation at these distances. It is not possible to distinguish between other distant activity mechanisms without denser heliocentric light curve data of higher precision, including observations at larger distances.

\begin{figure*}[ht!]
\plotone{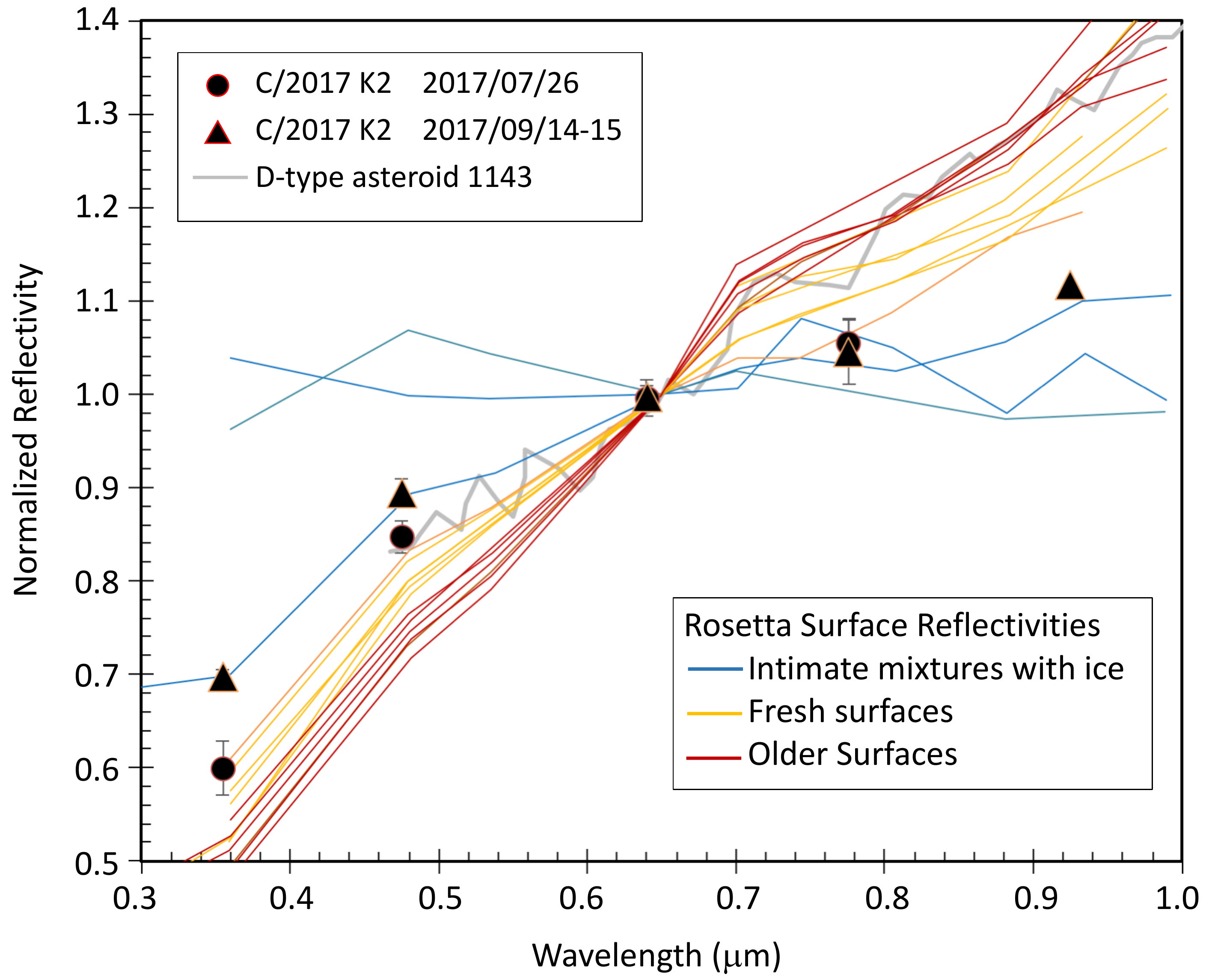}
\caption{Spectral reflectivity of C/2017 K2 obtained on 2017 July 26 and 2017 September 14-15 (see Table~\ref{tab:data}) compared to reflectivities of different surface types on C/67P Churyumov-Gerasimenko from the OSIRIS imaging system \citep{fornasier2017}.  }
\label{fig:spectrum}
\end{figure*}

The comet's spectral reflectivity falls within the envelope of the different regions on comet 67P (Fig.~\ref{fig:spectrum}).  Many regions on comet 67P were similar to or redder than typical D-type asteroids, and were dominated by organic-rich refractory material.  It was observed that 67P became spectrally less red overall as it approached perihelion and dust was removed, exposing underlying water ice \citep{fornasier2017}.  The spatially resolved spectral reflectivities show that newly exposed materials were less red, while in regions with spectroscopic signature of water-ice frost, the spectrum became progressively bluer, with the flat reflectivities having 20-32\% water-ice frost.  The reflectivity slope of C/2017 K2 is more consistent with 67P surfaces that contained some water-ice frost.  This could be the result of strong sublimation from near-surface CO for many years.

\begin{figure*}[ht!]
\plotone{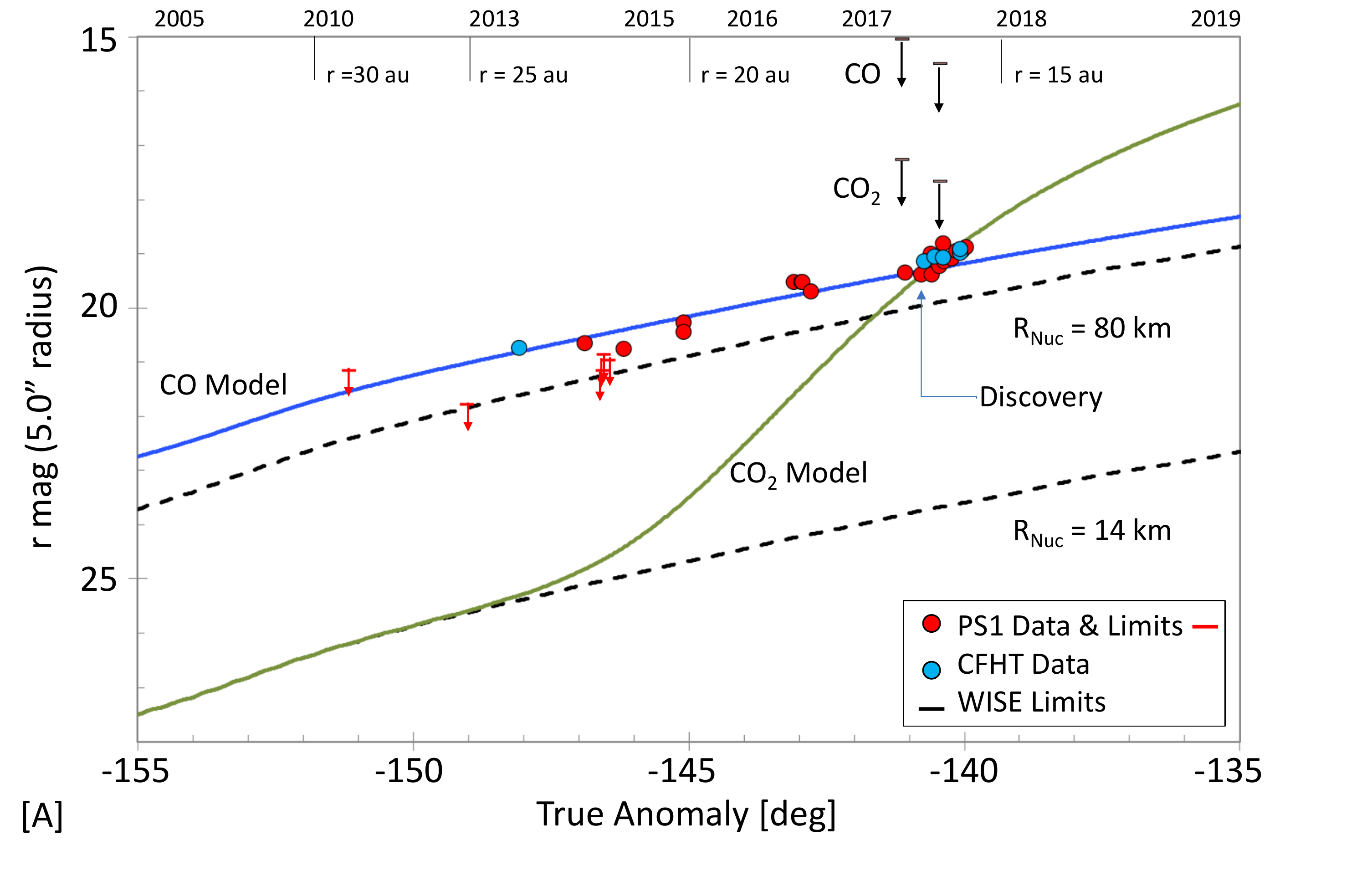}
\plotone{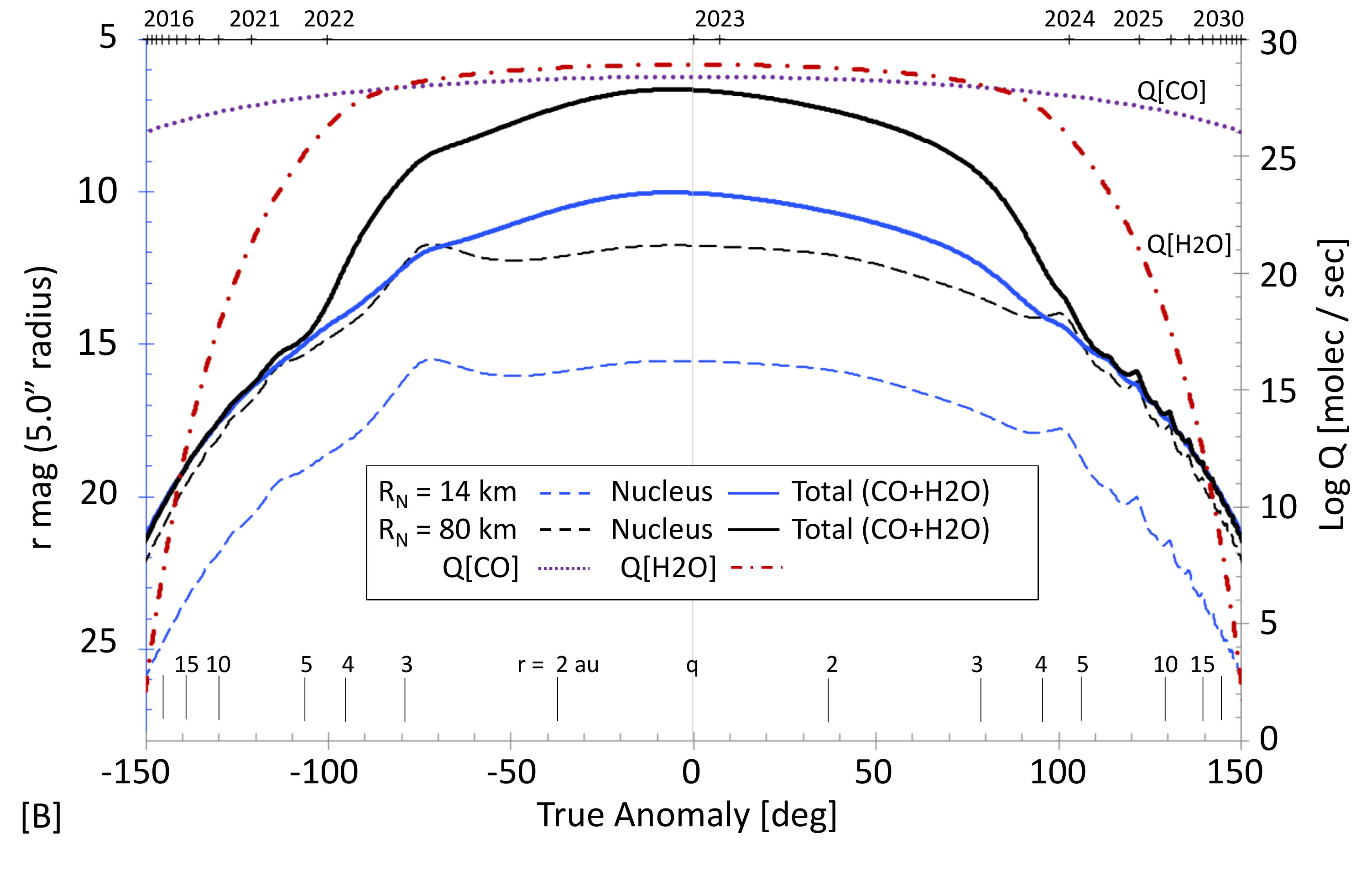}
\caption{[A] Best fit CO-sublimation model (blue) compared to the photometry in Table~\ref{tab:data} for our two limiting nucleus sizes. A CO$_2$ model (green) for a 14 km-radius nucleus consistent with the photometry at the time of discovery is also shown, but is inconsistent with the earlier PS1 data.  The upper limiting model magnitudes that correspond to the flux limits from the WISE observatory are also shown.
[B] Run out of the CO-sublimation models for 2 nucleus sizes assuming constant sublimating area for both H$_2$O and CO.  Under this assumption, and if there are no seasonal effects, this plot can be used to estimate the comet's brightness and H$_2$O and CO production rates.  These rates are shown for the $R_N$ = 14 km nucleus (dotted and dash-dot lines). For the $R_N$ = 80 km nucleus case the $Q_{CO}$ curve shifts by -0.37 and the $Q_{H2O}$ curve shifts by +1.51 in log.
\label{fig:model}}
\end{figure*}

In order to provide some guidance to observers who may want to plan observing runs to watch the development of activie, in Fig.~\ref{fig:model}B we run the models for both limiting nucleus cases through perihelion out to 25 au post-perihelion.  On the assumption that the fractional active areas of CO and H$_2$O do not change and there are no seasonal effects, the peak brightness of the comet should be between magnitude 7-11 through a 5$''$ aperture.  

On the right side of Fig.~\ref{fig:model}B we show the corresponding estimated production rates for both volatiles. To estimate the detectability of volatile species and D/H isotopologues at infrared wavelengths by ground-based observatories, we use a Figure of Merit (FoM). FoM is used to gauge the strength of molecular line emission. Traditionally, FoM = 10$^{29}$\,$\times$\,Q\,$\times$r$^{-1.5}$$\times$$\Delta$$^{-1}$, where Q is the H$_2$O production rate (molecules\,s$^{-1}$) predicted by our models, and $r$ and $\Delta$ are heliocentric and geocentric distance, in au. Typically, for a comet with FoM $\geq$ 0.08 we expect to measure H$_2$O and for FoM $\geq$ 2 we expect to unambiguously detect HDO. Adopting the H$_2$O production rates predicted for the lower and upper $R_N$ limits of 14\,km and 80\,km, the FoM predicts that H$_2$O is detectable inside $r$$\sim$2.1\,au and $\sim$3.4\,au, respectively. If $R_N$ is 80\,km, then a D/H measurement would be possible inside $r$$\sim$2.1\,au (i.e. from about early September 2022 through late March 2023). Of course, it is highly likely that the mixing ratios will not remain constant; this is just a guide for planning observations.

The James Webb Space Telescope ($JWST$) will facilitate high SNR spectra in the 2--5$\mu$m region to characterize the chemical composition of comets through the resolved spectral signatures of H$_2$O, CO, and CO$_2$. However, $JWST$ pointing limitations restrict observations to solar elongations  between 85$^\circ$-135$^\circ$, limiting the windows of observability. According to the current $JWST$ launch window estimates, the earliest we can observe the comet will be at $\sim$\,11\,au. Adopting our model predicted CO production rate of $\sim$\,10$^{26}$ molecules\,sec$^{-1}$ at r\,$\sim$\,11\,au, one hour of on-source integration yields a spectrum with a S/N$\sim$50 across the CO$_2$ and CO wavelength region using NIRSpec with a medium resolution G395M filter. $JWST$ observations at this distance would provide the first fully resolved medium resolution spectral signatures of CO and CO$_2$ fundamental vibration bands in a pre-perihelion comet beyond 6.2\,au.

\begin{figure*}[ht!]
\plotone{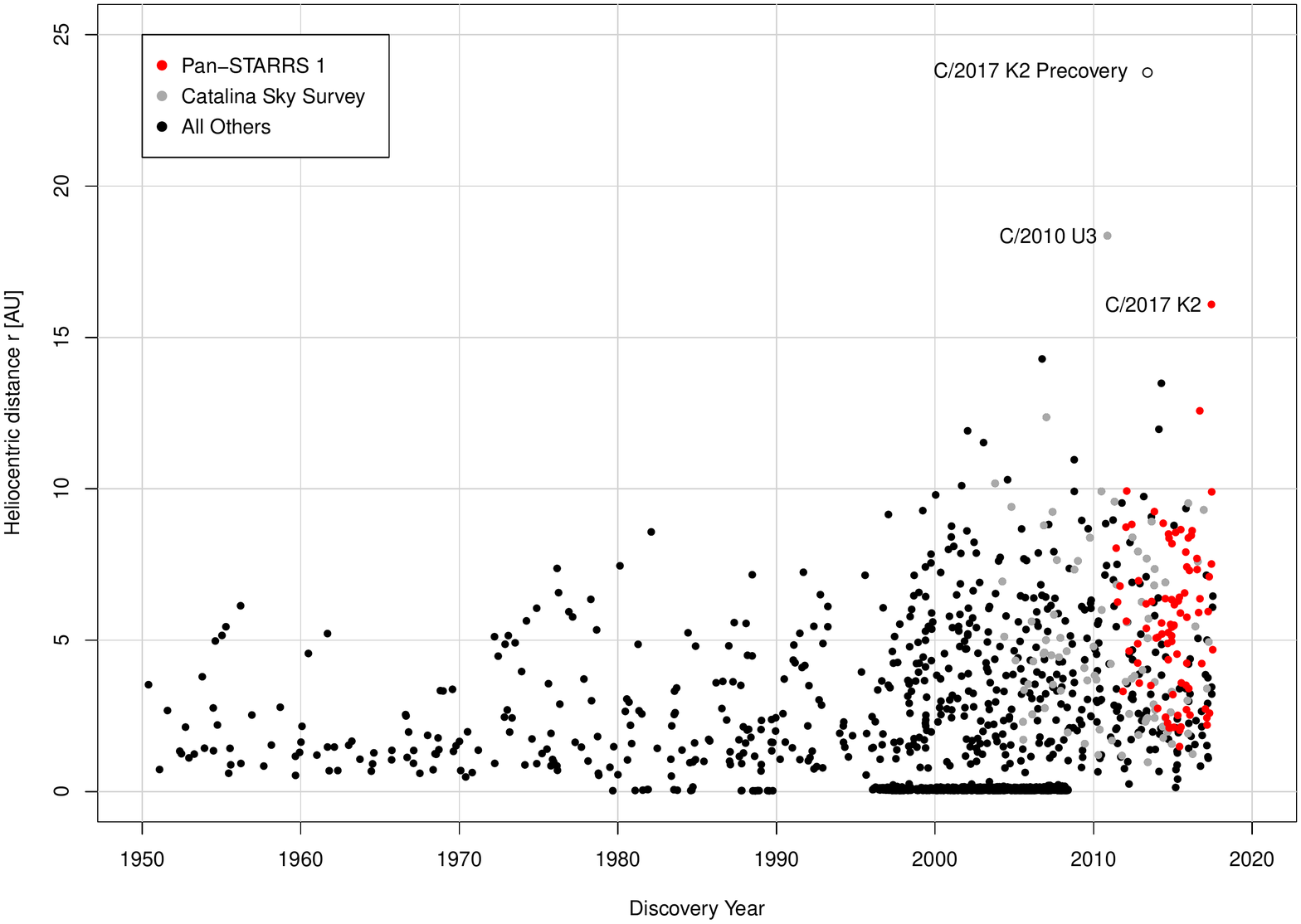}
\caption{Heliocentric distances at time of discovery of 2096 long-period comets discovered after 1950.  The points near 0\,au starting in 1996 are SOHO sun-grazing or impacting comets. The density of points increased significantly after 1996 when many of the major moving object sky surveys began.}
\label{fig:discovery}
\end{figure*}

Figure~\ref{fig:discovery} shows that all-sky surveys are finding more LP comets, and at larger distances.  Since 2010, of the $\sim$300 LP comets discovered, PS1 (31.3\%; shown as the red dots in Fig.~\ref{fig:discovery}) and CSS (26.1\%) are dominating the discoveries. Surprisingly, no survey or group is yet dominant for $r>10$ au. Since 2000 there have been 13 comets discovered beyond this distance. While some are discovered by surveys (PS1, Catalina, LONEOS, NEAT), others are discovered in deep targeted searches for distant trans-Neptunian objects.  The Pan-STARRS2 telescope \citep{morgan2012} will double the survey power of PS1 beginning in 2018. When the Large Synoptic Survey Telescope (LSST) begins its survey in 2023, we expect an explosion in distant LP comet discoveries that will enable a new understanding of cometary physics. With these surveys we may finally obtain observational confirmation of the activity that was predicted for historical comets as far out as 30 au \citep{sekanina1975}.


{\it Acknowledgements}
KJM, JTK, and JVK acknowledge support through awards from the National
Science Foundation AST1413736 and AST1617015. RJW acknowledges
support by the National Aeronautics and Space Administration under grant NNX14AM74G issued through the SSO Near Earth Object Observations Program. LD acknowledges support by NASA under grants NNX12AR55G and NNX14AM74G.

Based also in part on observations obtained with MegaPrime/MegaCam, a joint project of CFHT and CEA/DAPNIA, at the Canada-France-Hawaii Telescope (CFHT) which is operated by the National Research Council (NRC) of Canada, the Institute National des Science de l'Univers of the Centre National de la Recherche Scientifique (CNRS) of France, and the University of Hawai'i. 
This research used the facilities of the Canadian Astronomy Data Centre operated by the National Research Council of Canada with the support of the Canadian Space Agency.  

{\it Note added in Proof} -- During review, a paper by \citet{jewitt2017} on C/2017 K2 was published. Our nucleus radius lower limit is not in disagreement with the upper limit they presented.  Ours is a spherical equivalent radius, and their estimate is from an instantaneous measurement.  Nucleus axis ratios have been seen as high as 3.3 from the EPOXI mission.  The technique of using high-resolution HST measurements and coma removal typically produces agreement within $\pm$10-50\% of the spherical equivalent radius.

\begin{longrotatetable}
\begin{deluxetable*}{lcccccccccl}
\tablecaption{Observing Geometry and Photometry \label{tab:data}}
\tablehead{
\colhead{UTDate                    } &
\colhead{JD\tablenotemark{a}       } &
\colhead{$r$\tablenotemark{b}      } &
\colhead{$\Delta$\tablenotemark{b} } &
\colhead{$\alpha$\tablenotemark{b} } &
\colhead{TA\tablenotemark{c}       } &
\colhead{Filt                      } &
\colhead{\# Images                 } &
\colhead{mag$\pm$$\sigma$          } &
\colhead{r$_{mag}$$\pm$$\sigma$\tablenotemark{d}} &
\colhead{Color/Comment}
}
\startdata
\multicolumn{11}{l}{\bf PanSTARRS1 data}\\
2010/06/04 & 5352.10123	& 28.665 & 28.754 & 2.017 & -151.16 & i$_{p1}$ &  2 & $<$21.0$\pm$0.3   & $<$21.2$\pm$0.3  & Limiting mag \\
2012/08/07 & 6146.81325	& 25.061 & 25.146 & 2.306 & -149.01 & r$_{p1}$ &  2 & $<$21.8$\pm$0.3   & $<$21.8$\pm$0.3  & Limiting mag \\
2014/06/14 & 6821.46247 & 21.802 & 21.829 & 2.666 & -146.54 & i$_{p1}$ &  3 & $<$21.0$\pm$0.3   & $<$21.2$\pm$0.3  & Limiting mag \\ 
2014/06/13 & 6822.51886 & 21.797 & 21.824 & 2.667 & -146.54 & i$_{p1}$ &  3 & $<$20.7$\pm$0.3   & $<$20.9$\pm$0.3  & Limiting mag \\
2014/07/05 & 6844.27742 & 21.688 & 21.729 & 2.681 & -146.45 & i$_{p1}$ &  2 & $<$20.8$\pm$0.3   & $<$21.0$\pm$0.3  & Limiting mag \\ 
2014/03/20 & 6737.07683 & 22.221 & 22.202 & 2.569 & -146.88 & r$_{p1}$ &  2 &    20.67$\pm$0.30 & 20.67$\pm$0.30 & \\
2014/09/08 & 6908.76850 & 21.364 & 21.430 & 2.692 & -146.17 & i$_{p1}$ &  4 &    20.60$\pm$0.05 & 20.77$\pm$0.05 & \\
2015/05/06 & 7149.01152 & 20.138 & 20.122 & 2.871 & -145.09 & i$_{p1}$ &  4 &    20.10$\pm$0.04 & 20.28$\pm$0.04 & \\
2015/05/07 & 7150.00904 & 20.133 & 20.117 & 2.872 & -145.08 & i$_{p1}$ &  4 &    20.27$\pm$0.05 & 20.45$\pm$0.05 & \\
2016/05/27 & 7536.02124 & 18.088 & 18.052 & 3.211 & -143.07 & r$_{p1}$ &  4 &    19.53$\pm$0.04 & 19.53$\pm$0.04 & \\
2016/06/19 & 7558.88673 & 17.963 & 17.938 & 3.242 & -142.94 & i$_{p1}$ &  3 &    19.37$\pm$0.08 & 19.54$\pm$0.08 & \\
2016/06/21 & 7560.93624 & 17.952 & 17.928 & 3.245 & -142.93 & i$_{p1}$ &  2 &    19.37$\pm$0.06 & 19.54$\pm$0.06 & \\
2016/07/18 & 7587.88946 & 17.805 & 17.803 & 3.270 & -142.77 & i$_{p1}$ &  4 &    19.53$\pm$0.05 & 19.71$\pm$0.05 & \\
2017/04/10 & 7854.05173 & 16.322 & 16.280 & 3.519 & -141.06 & i$_{p1}$ &  4 &    19.19$\pm$0.07 & 19.36$\pm$0.07 & \\
2017/05/21 & 7894.91140 & 16.089 & 16.019 & 3.604 & -140.77 & w$_{p1}$ &  4 &    19.45$\pm$0.03 & 19.39$\pm$0.03 & Discovery \\
2017/06/16 & 7920.97821 & 15.940 & 15.874 & 3.651 & -140.59 & i$_{p1}$ &  4 &    18.85$\pm$0.04 & 19.02$\pm$0.04 & \\
2017/06/25 & 7929.91675 & 15.888 & 15.828 & 3.666 & -140.52 & w$_{p1}$ &  4 &    19.19$\pm$0.03 & 19.13$\pm$0.03 & \\
2017/08/06 & 7971.88255 & 15.646 & 15.638 & 3.715 & -140.21 & i$_{p1}$ &  4 &    18.94$\pm$0.05 & 19.11$\pm$0.05 & \\
2017/08/17 & 7982.77565 & 15.583 & 15.594 & 3.721 & -140.13 & w$_{p1}$ &  4 &    19.02$\pm$0.01 & 18.97$\pm$0.02 & \\
2017/08/30 & 7995.76736 & 15.508 & 15.542 & 3.725 & -140.03 & i$_{p1}$ &  4 &    18.78$\pm$0.04 & 18.96$\pm$0.04 & \\
2017/09/07 & 8003.77876 & 15.461 & 15.511 & 3.724 & -139.96 & i$_{p1}$ &  4 &    18.72$\pm$0.04 & 18.90$\pm$0.04 & \\
2017/06/16 & 7920.95557 & 15.940 & 15.874 & 3.651 & -140.59 & r$_{p1}$ &  4 &    18.96$\pm$0.06 & 19.14$\pm$0.06 & \\
2017/07/05 & 7939.88379 & 15.831 & 15.779 & 3.681 & -140.45 & r$_{p1}$ &  2 &    18.94$\pm$0.14 & 19.11$\pm$0.14 & \\
2017/07/07 & 7941.90196 & 15.819 & 15.770 & 3.684 & -140.43 & r$_{p1}$ &  4 &    19.24$\pm$0.11 & 19.41$\pm$0.11 & \\
2017/07/15 & 7949.87944 & 15.773 & 15.733 & 3.694 & -140.37 & r$_{p1}$ &  4 &    18.82$\pm$0.06 & 19.00$\pm$0.06 & \\
2017/07/17 & 7951.90526 & 15.762 & 15.724 & 3.696 & -140.36 & r$_{p1}$ &  4 &    19.15$\pm$0.20 & 19.33$\pm$0.20 & \\
\hline
\multicolumn{11}{l}{\bf CFHT Archival data from CADC} \\
2013/05/12 & 6424.61132 & 23.744 & 23.767 & 2.436 & -148.07 & u        &  7 &    23.09$\pm$0.17 & 20.76$\pm$0.23 & \\
\hline
\multicolumn{10}{l}{\bf CFHT new data} \\
2017/05/24 & 7898.05595 & 16.071 & 16.000 & 3.610 & -140.75 & w        &  1 &    19.292$\pm$0.007 & 19.237$\pm$0.007 & \\
2017/05/28 & 7901.92345 & 16.049 & 15.978 & 3.617 & -140.72 & g        &  1 &    19.673$\pm$0.015 & & \\
2017/05/28 & 7901.92345 & 16.049 & 15.978 & 3.617 & -140.72 & r        &  1 &    19.155$\pm$0.015 & 19.155$\pm$0.015 & (g-r) = 0.52$\pm$0.02 \\
2017/06/24 & 7928.97060 & 15.894 & 15.833 & 3.665 & -140.53 & g        &  1 &    19.539$\pm$0.012 & & \\
2017/06/24 & 7928.97049 & 15.894 & 15.833 & 3.665 & -140.53 & r        &  1 &    19.067$\pm$0.013 & 19.067$\pm$0.013 & (g-r) = 0.47$\pm$0.02 \\
2017/07/16 & 7950.83854 & 15.768 & 15.729 & 3.695 & -140.37 & r        & 12 &    19.081$\pm$0.004 & 19.081$\pm$0.004 & \\
2017/07/25 & 7990.80063 & 15.768 & 15.729 & 3.695 & -140.37 & g        &  1 &    19.567$\pm$0.011 & & (g-r) = 0.56$\pm$0.02 \\
2017/07/25 & 7990.80285 & 15.768 & 15.729 & 3.695 & -140.37 & r        &  1 &    19.007$\pm$0.012 & 19.007$\pm$0.012 & \\
2017/07/26 & 7991.80022 & 15.537 & 15.562 & 3.724 & -140.06 & u        &  2 &    21.260$\pm$0.054 & & (u-r) = 2.33$\pm$0.06 \\
2017/07/26 & 7991.80602 & 15.531 & 15.558 & 3.724 & -140.06 & g        &  1 &    19.513$\pm$0.013 & & (g-r) = 0.59$\pm$0.02 \\
2017/07/26 & 7991.80407 & 15.537 & 15.562 & 3.724 & -140.06 & r        &  3 &    18.927$\pm$0.008 & 18.927$\pm$0.008 & \\
2017/07/26 & 7991.80889 & 15.531 & 15.558 & 3.724 & -140.06 & i        &  2 &    18.758$\pm$0.013 & & (r-i)\hspace{0.1cm} = 0.17$\pm$0.02 \\
2017/09/14 & 8010.74649	& 15.421 & 15.483 & 3.723 & -139.91 & g	       &  4 &    19.483$\pm$0.005 & & (g-r) = 0.55$\pm$0.01 \\
2017/09/14 & 8010.74282	& 15.421 & 15.483 & 3.723 & -139.91 & r	       &  2 &    18.932$\pm$0.009 & 18.932$\pm$0.009 & \\
2017/09/14 & 8010.74230	& 15.421 & 15.483 & 3.723 & -139.91 & i	       &  6 &    18.761$\pm$0.008 & & (r-i)\hspace{0.1cm} = 0.17$\pm$0.01 \\
2017/09/14 & 8010.75793	& 15.421 & 15.483 & 3.723 & -139.91 & z	       &  4 &    18.721$\pm$0.024 & & (r-z) = 0.21$\pm$0.03 \\
2017/09/15 & 8011.72479	& 15.415 & 15.479 & 3.723 & -139.90 & u	       &  3 &    20.995$\pm$0.039 & & (u-r) = 2.12$\pm$0.04 \\
2017/09/15 & 8011.74031	& 15.415 & 15.479 & 3.723 & -139.90 & g	       &  5 &    19.413$\pm$0.004 & & (g-r) = 0.54$\pm$0.01 \\
2017/09/15 & 8011.75514	& 15.415 & 15.479 & 3.723 & -139.90 & r	       &  2 &    18.873$\pm$0.007 & 18.873$\pm$0.007 & \\
\enddata
\tablenotetext{a}{Julian Date -2450000.0}
\tablenotetext{b}{Heliocentric, geocentric distance (au) and phase angle (deg).}
\tablenotetext{c}{True anomaly (deg), position along orbit, TA at perihelion=0$^{\circ}$}
\tablenotetext{d}{Magnitude and error through 5$''$ radius aperture.}
\tablenotetext{e}{Magnitude and error converted to SDSS $r$ as described in the text}
\end{deluxetable*}
\end{longrotatetable}

\end{document}